\documentstyle[aps,floats,prl,epsfig]{revtex}

\begin{document}

\twocolumn[\hsize\textwidth\columnwidth\hsize\csname @twocolumnfalse\endcsname

\title{Tight-binding study of high-pressure phase transitions in
titanium:  alpha to omega and beyond}

\author{M. J. Mehl and D. A. Papaconstantopoulos}

\address{Center for Computational Materials Science, Naval
Research Laboratory, Washington DC 20375-5000}

\date{\today}

\maketitle

\begin{abstract}
We use a tight-binding total energy method, with parameters
determined from a fit to first-principles calculations, to examine
the newly discovered $\gamma$ phase of titanium.  Our parameters
were adjusted to accurately describe the $\alpha$Ti-$\omega$Ti phase
transition, which is misplaced by density functional calculations.
We find a transition from $\omega$Ti to $\gamma$Ti at 102~GPa, in
good agreement with the experimental value of 116~GPa.  Our results
suggest that current density functional calculations will not
reproduce the $\omega$Ti-$\gamma$Ti phase transition, but will
instead predict a transition from $\omega$Ti to the bcc $\beta$Ti
phase.
\end{abstract}

]

Structural transformations in titanium have received a great deal of
experimental
\cite{vohra01:novelti,sikka82:omega_phase,dobromyslov90:tiomega,xia90:hf}
and
theoretical\cite{jomard98:ti_gga,gyanch90:tigga,ahuja93:tizrhf,greeff01:ti_shock,nishitani01:ti:bcc-hcp,ostanin97:ti}
attention.  This Letter is motivated by a recent experimental
study\cite{vohra01:novelti} which revealed a previously unsuspected
phase transition in titanium at 116~GPa from the $\omega$Ti phase to
a new $\gamma$Ti phase.  We have been able to confirm these
experiments by performing highly accurate tight-binding calculations
of the phase diagram of Ti.

At room temperature the group-IV metals zirconium and hafnium
transform under pressure from the ground state hexagonal close
packed (hcp) phase to the intermediate pressure $\omega$
phase\cite{sikka82:omega_phase} (space group $P6/mmm - D_{6h}^1$,
Pearson Symbol hP3, {\em Strukturbericht} Designation: C32) at
2.2~GPa\cite{sikka82:omega_phase} and 38~GPa\cite{xia90:hf},
respectively.  At 35~GPa\cite{xia90:bcczr,xia91:omega_bcc} and
71~GPa\cite{xia90:hf}, respectively, the metals transform from the
$\omega$ phase to a body centered cubic (bcc) structure.  One would
logically assume that titanium also follows this transformation
sequence, and indeed the transition from hcp $\alpha$Ti to
$\omega$Ti takes place at a pressure variously given as
9~GPa\cite{vohra01:novelti} or 20-90~GPa\cite{sikka82:omega_phase}.
However, no room temperature pressure driven transition from
$\omega$Ti to $\beta$Ti has been observed, although first-principles
calculations predict a transition at
98~GPa,\cite{jomard98:ti_gga,ostanin97:ti} and $\beta$Ti exists at
room pressure and temperatures above
1155~K.\cite{massalski87:binary} Recently, however, Vohra and
Spencer\cite{vohra01:novelti} found that at 116~GPa titanium
transforms from the $\omega$ phase to a previously unsuspected
$\gamma$Ti phase.  The new phase has a two-atom body-centered
orthorhombic unit cell, space group $Cmcm$--$D_{2h}^{17}$, Pearson
symbol oC4, with the atoms at the points $( 0 , \pm y b , \pm c/4
)$, where $a$, $b$, and $c$ are the lengths of the primitive vectors
in the full orthorhombic unit cell, and $y$ is an internal
parameter.

This crystal structure is observed in various materials, including
$\alpha$U, the random alloy AgCd, and a metastable form of
gallium.\cite{pearson_handbook} (The $\alpha$U phase has the {\em
Strukturbericht} designation A20.\cite{strukture,note:a20}) With
appropriate choices of the parameters this structure can reproduce
several higher symmetry phases.  In particular, when $b/a = \sqrt3$
and $y = 1/6$, it becomes the hcp structure, while when $b/a = c/a =
\sqrt2$ and $y = 1/4$ the atoms are on the sites of a bcc cell.
Wentzcovitch and Cohen\cite{wentzcovitch88:_mg:hcp_to_bcc} used this
pathway to describe a possible theoretical model for the hcp
$\rightarrow$ bcc transition in magnesium.

Examination of the $\gamma$Ti structure by first-principles
techniques requires a minimization of the total energy with respect
to three parameters ({\em e.g.}, $b/a$, $c/a$, and $y$) at several
volumes.  This is impractical because of the high computational
demand of first principles methods.  We have instead chosen to study
the $\alpha$-$\omega$-$\gamma$ transformation sequence using the
much faster NRL tight-binding
method.\cite{cohen94:_tight,mehl96:_appli} This method has been
shown to reproduce the ground-state phase, elastic constants,
surface energies, and vacancy formation energies of the transition
metals.  The tight-binding parameters in Ref.~\cite{mehl96:_appli}
were found by fitting to a Local Density Approximation (LDA)
database of total energies and eigenvalues for the fcc and bcc
structures.  The parameters correctly predicted the ground state
structures of all of the transition and noble metals, including the
hcp metals and manganese.\cite{mehl95:tb_mn} However, upon
examination, we found that the titanium parameters from
Ref.~\cite{mehl96:_appli} do not predict the correct position for
the $\omega$Ti phase.  In fact, no $\alpha$Ti-$\omega$Ti phase
transition is seen.

We therefore developed a new set of tight-binding parameters
according to the procedures of Ref.~\cite{mehl96:_appli}, fit to an
expanded database of first-principles calculations.  In particular,
our database includes the fcc, bcc, simple cubic, hcp, and $\omega$
structures.  The eigenvalues and total energies were generated using
the general-potential Linearized Augmented Plane Wave (LAPW)
method,\cite{andersen75:_linea,wei85:lapw} using the Perdew-Wang
1991 Generalized Gradient Approximation
(GGA)\cite{perdew91,perdew92:gga_apps} to density functional theory.
We fit our tight-binding parameters to both total energies and band
structures, using the parametrization described by equations (7),
(8), (9), and (11) of Ref.~\cite{mehl96:_appli}.  The RMS error in
fitting the energies for the lowest energy phases (hcp, $\omega$,
fcc, and bcc) was less than 1~mRy/atom.  The band structure RMS
error is about 10~mRy for the occupied bands of the hcp and $\omega$
structures.\cite{note:tipar}

\begin{table}
\caption{The equilibrium lattice constants of the $\alpha$ (hcp),
$\beta$ (bcc), and $\omega$ phases of titanium, as determined by the
tight-binding parameters described in the
text,\protect\cite{note:tipar} the LAPW calculations used in the
fitting procedure, and experiment.  Note that $\beta$Ti is not seen
at room temperature.  The lattice constant given is extrapolated
from alloy data.\protect\cite{donohue74:elements} All values are in
Bohr.}
\begin{tabular}{ccccccc}
Phase & \multicolumn{2}{c}{TB} & \multicolumn{2}{c}{LAPW} &
\multicolumn{2}{c}{Exp.} \\
& a & c & a & c & a & c \\
\tableline
$\alpha$\tablenotemark[1] & \dec 5.561 & \dec 8.609 & \dec 5.547 &
\dec 8.779 & \dec 5.575 & \dec 8.851 \\
$\beta$\tablenotemark[1] & \dec 6.118 & \dec 6.118 & \dec 6.137 &
\dec 6.137 & \dec 6.206 & \dec 6.206 \\
$\omega$\tablenotemark[2] & \dec 8.675 & \dec 5.268 & \dec 8.644 &
\dec 5.348 & \dec 8.689 & \dec 5.333 \\
\end{tabular}
\tablenotetext[1]{Experimental data from
Ref.~\cite{donohue74:elements}}
\tablenotetext[2]{Experimental data from
Ref.~\cite{vohra01:novelti}}
\label{tab:eqlat}
\end{table}

In agreement with previous
calculations,\cite{jomard98:ti_gga,gyanch90:tigga,ahuja93:tizrhf,greeff01:ti_shock}
our first-principles results show that the $\omega$Ti phase is
slightly lower in energy (about 0.5~mRy/atom) than $\alpha$Ti.  This
implies a $-5$~GPa $\omega$Ti-$\alpha$Ti phase transition.  We have
adjusted our tight binding parametrization to shift the $\omega$Ti
phase equilibrium upwards by 0.8~mRy/atom.  This produces an
$\alpha$Ti-$\omega$Ti phase transition at 6~GPa, in good agreement
with the 9~GPa transition found in Ref.~\cite{vohra01:novelti} As we
shall see, this has important consequences for the
$\omega$Ti-$\gamma$Ti phase transition.

\begin{figure}
\epsfig{file=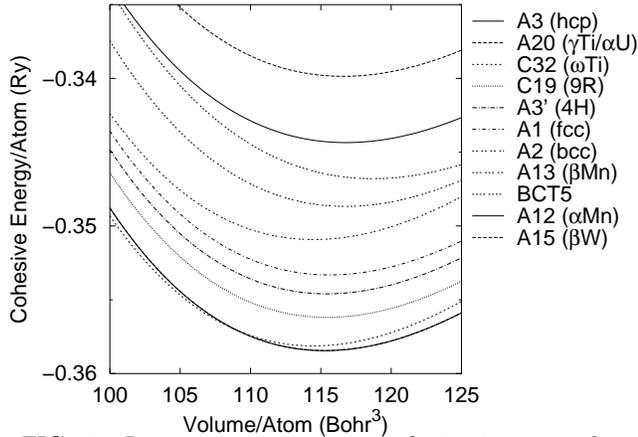,width=3.4in}
\caption{Low energy structures of titanium, as determined by the
tight-binding parameters described in the text.  The $\gamma$Ti
phase is described in the text.  Over this range of volumes it is
degenerate with the hcp (A3, or $\alpha$Ti) structure.  The crystal
structures are described in full at
http://cst-www.nrl.navy.mil/lattice.}
\label{fig:loweos}
\end{figure}

We have tested the fit in a variety of ways.  In
Table~\ref{tab:eqlat} we show the equilibrium lattice constants for
$\alpha$-, $\beta$- and $\omega$Ti, as determined by our
tight-binding parameters, our first-principles LAPW calculations,
and experiment.\cite{vohra01:novelti,donohue74:elements} The TB
agreement with experiment is comparable to that achieved by the
first-principles calculations.

We also examined the behavior of a variety of crystal structures
using our tight-binding parameters.  Fig.~\ref{fig:loweos} shows the
energy/volume behavior of a number of low energy structures.  As
expected, the observed phases ($\alpha$Ti and $\omega$Ti) are
followed by close-packed stacking fault phases (9R, 4H, and fcc),
and then other common metallic phases.

We further checked the behavior of our tight-binding Hamiltonian by
determining the elastic constants and phonon frequencies in
$\alpha$Ti, as shown in Tables~\ref{tab:cij} and \ref{tab:phonon},
respectively.  Compared to experiment, we find an RMS error of
22~GPa for the elastic constants, and 32~cm$^{-1}$ for the phonon
frequencies.  This is typical of the predictive capability of the
tight-binding method for hcp metals.\cite{mehl96:_appli}

\begin{table}
\caption{Elastic constants of $\alpha$Ti at the experimental volume,
as determined from the parameters described in the text and compared
to experiment.\protect\cite{simmons71:cij} All values are in GPa.}
\begin{tabular}{rrr|rrr}
 & \multicolumn{1}{c}{TB} & \multicolumn{1}{c}{Exp.} &  &
 \multicolumn{1}{c}{TB} & \multicolumn{1}{c}{Exp.} \\
\tableline
$C_{11}$ & 127 & 162 & $C_{33}$ & 147 & 181 \\
$C_{12}$ & 81 & 92 & $C_{44}$ & 45 & 47 \\
$C_{13}$ & 64 & 69 \\
\end{tabular}
\label{tab:cij}
\end{table}

\begin{table}
\caption{High-symmetry {\bf k}-point phonon frequencies (in
cm$^{-1}$) of $\alpha$Ti at the experimental volume, as determined
from the parameters described in the text and compared to
experiment.\protect\cite{stassis79:ti_phonons} The symmetry notation
is from Miller and
Love.\protect\cite{miller67:spcgrp,stokes88:spcgrp} }
\begin{tabular}{ccrr|ccrr}
 &  & TB & Exp. &  &  & TB & Exp. \\
\tableline
$\Gamma_3^+$ & & 189 & 185 \\
$\Gamma_5^+$ & & 141 & 137 & $M_1^+$ & LO & 259 & 257 \\
\cline{1-4}
$A_1$ & LA/LO & 178 & 191  & $M_2^+$ & TA & 56 & 113 \\ 
$A_3$ & TA/TO & 98 & 101 & $M_2^-$ & TO & 171 & 202 \\
\cline{1-4}
$K_1$ & & 275 & 234 & $M_3^+$ & TA & 99 & 127 \\
$K_4$ & & 180 & 200 & $M_3^-$ & TO & 192 & 232 \\
$K_5$ & LA/LO & 144 & 207 & $M_4^-$ & LA & 200 & 234 \\
$K_6$ & TA/TO & 141 & 173 \\
\end{tabular}
\label{tab:phonon}
\end{table}

\begin{figure}
\epsfig{file=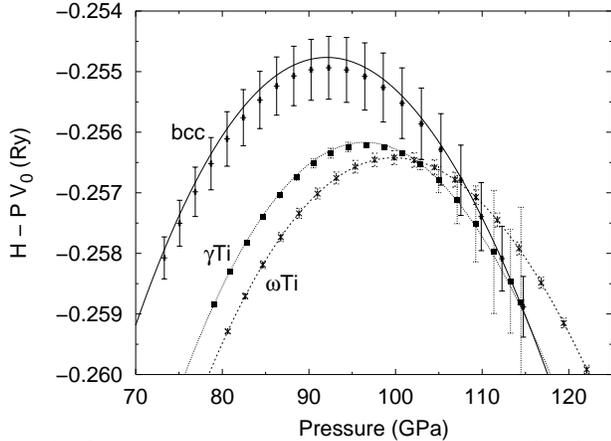,width=3.4in}
\caption{The enthalpy for several phases of titanium in the
$\omega$-$\gamma$ transition region.  For increased clarity we
subtract $P V_0$ from each enthalpy, where the reference volume $V_0
= 75$~Bohr$^3$.  The lines are derived from extended Birch
fits\protect\cite{birch78:eos,mehl94:fpcij} to the $E(V)$ data for
each phase.  Numerical differentiation of $E(V)$ provides the
pressure for the points, and a check on the accuracy of the Birch
fit.  The error bars on the points represent estimates of the
uncertainty in the energy and pressure calculations due to the
numerical k-point integration.}
\label{fig:trans}
\end{figure}

We studied the $\alpha$-$\omega$-$\gamma$ transition path in
titanium by fixing the volume of a given phase, and then minimizing
the total energy as a function of the other parameters ($c/a$ for
$\alpha$ (hcp) and $\omega$; $b/a$, $c/a$, and $y$ for $\gamma$).
The pressure was calculated in one of two ways: by differentiating
an extended Birch fit,\cite{birch78:eos,mehl94:fpcij} and by
calculating the pressure by numerical differentiation of the total
energy with respect to volume.  The enthalpy of each phase, $H(P) =
E + P V$ is then calculated by both methods.  In
Fig.~\ref{fig:trans} we show the enthalpy of the $\omega$Ti,
$\gamma$Ti, and bcc ($\beta$Ti) phases in the transition region.
From the plot we see that the $\omega$Ti-$\gamma$Ti phase transition
takes place at about 102~GPa, compared to the experimental result of
116~GPa.  We also see a $\gamma$Ti-$\beta$Ti phase transition at
about 115~GPa.  This is not seen experimentally, but it suggests
that we may expect a higher pressure $\gamma$Ti-$\beta$Ti phase
transition, which would complete the $\alpha$-$\omega$-$\beta$
transition sequence seen in Zr and Hf, albeit with an interloping
$\gamma$Ti phase between $\omega$Ti and $\beta$Ti.  More details of
the phase transitions predicted by our Hamiltonian are shown in
Table~\ref{tab:trans}.

In the absence of the $\gamma$Ti phase, Fig.~\ref{fig:trans} shows
that there would be an $\omega$Ti-$\beta$Ti phase transition at
110~GPa.  This is in good agreement with the prediction made from
the LAPW/GGA calculations in our database, 105~GPa, and with the
LMTO/GGA prediction of 98~GPa found in Ref.~\cite{jomard98:ti_gga}.

\begin{table}
\caption{Pressure induced phase transitions in titanium, as
determined by the tight-binding parameters described in the
text\protect\cite{note:tipar} and compared to
experiment.\protect\cite{vohra01:novelti}.}
\begin{tabular}{ccccc}
 & \multicolumn{2}{c}{TB} & \multicolumn{2}{c}{Exp.} \\
Transition & Pressure & $\Delta V/V$ & Pressure & $\Delta V/V$ \\
 & (GPa) & (\%) & (GPa) & (\%) \\
\tableline
$\alpha \rightarrow \omega$ & 6 & -0.6 & 9 & -1.9 \\
$\omega \rightarrow \gamma$ & 102 & -1.3 & 116 & -1.6 \\
$\gamma \rightarrow \beta$ & 115 & $\approx$ 0 &
\multicolumn{2}{c}{none up to 146~GPa}
\end{tabular}
\label{tab:trans}
\end{table}

\begin{figure}
\epsfig{file=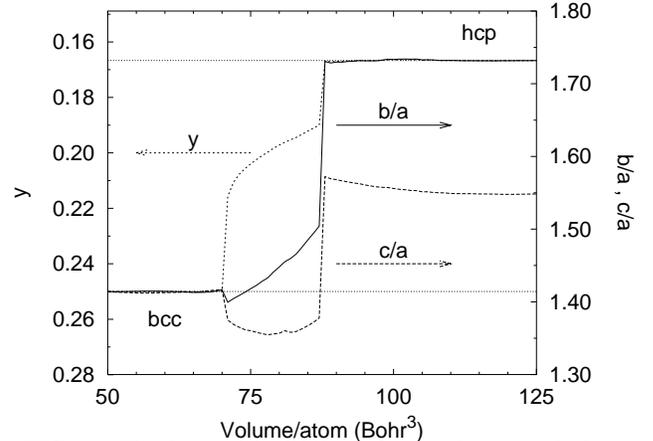,width=3.4in}
\caption{The lattice parameters $b/a$ and $c/a$ and the internal
parameter $y$ which minimize the total energy of $\gamma$Ti as a
function volume, using the tight-binding parameters described in the
text.  The horizontal dotted lines indicate the parameter values
needed to achieve an ideal bcc lattice ($y = 1/4$, $b/a = c/a =
\sqrt2$) and an hcp lattice ($y = 1/6$, $b/c = \sqrt3$, arbitrary
$c/a$).}
\label{fig:latpar}
\end{figure}

The behavior of titanium in the $\alpha$-, $\beta$-, and $\gamma$Ti
phases is explored further in Fig.~\ref{fig:latpar}, which shows the
lattice parameters $b/a$, $c/a$, and $y$ as a function of the
volume.  We see that at a volume of about 85~Bohr$^3$/atom there is
an abrupt change from hcp $\alpha$Ti into the lower symmetry
$\gamma$Ti phase.  From this point the structure merges more or less
continuously into bcc $\beta$Ti at about 70~Bohr$^3$/atom.  Note,
however, that none of these phases is observed in the volume range
74-108~Bohr$^3$, as this is the region where the $\omega$Ti phase is
stable.

Finally, we note that our LAPW calculations and other
first-principles
calculations\cite{jomard98:ti_gga,gyanch90:tigga,ahuja93:tizrhf,greeff01:ti_shock}
place the $\omega$Ti phase slightly lower in energy than the
$\alpha$Ti phase, leading to a direct transition from $\omega$Ti to
$\beta$Ti at 105~GPa.  Hence, the essential difference between the
first-principles calculations and our TB model is the ordering of
the $\alpha$Ti and $\omega$Ti phases.

In conclusion, our tight-binding Hamiltonian provides a good
description of the low pressure behavior of titanium, and shows the
correct $\alpha$-$\omega$-$\gamma$ transition sequence as reported
in recent experiments.  Our work suggests that current
first-principles density functional calculations, which place the
$\omega$Ti phase below the $\alpha$Ti phase, will also fail to
predict the stability of the $\gamma$Ti phase under pressure.

We thank I. I. Mazin for reminding us of
Ref.~\cite{wentzcovitch88:_mg:hcp_to_bcc}, and T. A. Adler for
pointing us to the $\alpha$U structure.  This work was supported by
the U. S. Office of Naval Research.  The development of the
tight-binding codes was supported in part by the U. S. Department of
Defense Common HPC Software Support Initiative (CHSSI).

\begin{thebibliography}{10}

\bibitem{vohra01:novelti}
Y.~K. Vohra and P.~T. Spencer, Phys. Rev. Lett. {\bf 86},  3068  (2001).

\bibitem{sikka82:omega_phase}
S.~K. Sikka, Y.~K. Vohra, and R. Chidambaram, Prog. Mat. Sci. {\bf 27},  245
  (1982).

\bibitem{dobromyslov90:tiomega}
A.~V. Dobromyslov and N.~I. Taluts, Physics of metals and metallography {\bf
  69},  98  (1990).

\bibitem{xia90:hf}
H. Xia, G. Parthasarathy, H. Luo, Y.~K. Vohra, and A.~L. Ruoff, Phys. Rev. B
  {\bf 42},  6736  (1990).

\bibitem{jomard98:ti_gga}
G. Jomard, L. Magaud, and A. Pasturel, Phil. Mag. B {\bf 77},  67  (1998).

\bibitem{gyanch90:tigga}
J.~S. Gyanchandani, S.~C. Gupta, S.~K. Sikka, and R. Chidambaram, J.~Phys.:
  Cond. Matt. {\bf 2},  301  (1990).

\bibitem{ahuja93:tizrhf}
R. Ahuja, J.~M. Wills, B. Johansson, and O. Eriksson, Phys. Rev. B {\bf 48},
  16269  (1993).

\bibitem{greeff01:ti_shock}
C.~W. Greeff, D.~R. Trinkle, and R.~C. Albers, J.~Appl. Phys. {\bf 90},  2221
  (2001).

\bibitem{nishitani01:ti:bcc-hcp}
S.~R. Nishitani, H.~K. H, and M. Aoki, Materials Science and Engineering A {\bf
  312},  77  (2001).

\bibitem{ostanin97:ti}
S.~A. Ostanin and V.~Y. Trubitsin, Journal of Physics: Condensed Matter {\bf
  9},  L491  (1997).

\bibitem{xia90:bcczr}
H. Xia, S.~J. Duclos, A.~L. Ruoff, and Y.~K. Vohra, Phys. Rev. Lett. {\bf 64},
  204  (1990).

\bibitem{xia91:omega_bcc}
H. Xia, A.~L. Ruoff, and Y.~K. Vohra, Phys. Rev. B {\bf 44},  10374  (1991).

\bibitem{massalski87:binary}
{\em Binary Alloy Phase Diagrams}, edited by T.~B. Massalski (American Society
  for Metals, Metals Park, Ohio, 1987).

\bibitem{pearson_handbook}
{\em Pearson's Handbook of Crystallographic Data for Intermetallic Phases},
  2$^{nd}$ ed., edited by P. Villars and L. Calvert (ASM International,
  Materials Park, Ohio, 1991).

\bibitem{strukture}
{\em Strukturbericht}, edited by K. Herrman (Akademische Verlagsgesellschaft
  Becker \& Erler, Leipzig, 1938), Vol.~VI.

\bibitem{note:a20}
More information about the $\alpha$U (A20) structure can be obtained at\\
  http://cst-www.nrl.navy.mil/lattice/struk/a20.html.

\bibitem{wentzcovitch88:_mg:hcp_to_bcc}
R.~M. Wentzcovitch and M.~L. Cohen, Phys. Rev. B {\bf 37},  5571  (1988).

\bibitem{cohen94:_tight}
R.~E. Cohen, M.~J. Mehl, and D.~A. Papaconstantopoulos, Phys. Rev. B {\bf 50},
  14694  (1994).

\bibitem{mehl96:_appli}
M.~J. Mehl and D.~A. Papaconstantopoulos, Phys. Rev. B {\bf 54},  4519  (1996).

\bibitem{mehl95:tb_mn}
M.~J. Mehl and D.~A. Papaconstantopoulos, Europhys. Lett. {\bf 31},  537
  (1995).

\bibitem{andersen75:_linea}
O.~K. Andersen, Phys. Rev. B {\bf 12},  3060  (1975).

\bibitem{wei85:lapw}
S.-H. Wei and H. Krakauer, Phys. Rev. Lett. {\bf 55},  1200  (1985).

\bibitem{perdew91}
J.~P. Perdew,  in {\em Electronic Structure of Solids '91}, edited by P.
  Ziesche and H. Eschrig (Akademie Verlag, Berlin, 1991).

\bibitem{perdew92:gga_apps}
J.~P. Perdew, J.~A. Chevary, S.~H. Vosko, , K.~A. Jackson, M.~R. Pederson,
  D.~J. Singh, and C. Fiolhais, Phys. Rev. B {\bf 46},  6671  (1992).

\bibitem{note:tipar}
The titanium parameters discussed in this Letter are available at
  http://cst-www.nrl.navy.mil/bind/ti.html.

\bibitem{donohue74:elements}
J. Donohue, {\em The Structures of the Elements} (John Wiley \& Sons, New York,
  1974).

\bibitem{simmons71:cij}
G. Simmons and H. Wang, {\em Single Crystal Elastic Constants and Calculated
  Aggregate Properties: A HANDBOOK}, 2$^{nd}$ ed. (M.I.T. Press, Cambridge,
  Massachusetts and London, 1971).

\bibitem{stassis79:ti_phonons}
C. Stassis, D. Arch, B.~N. Harmon, and N. Wakabayashi, Phys. Rev. B {\bf 19},
  181  (1979).

\bibitem{miller67:spcgrp}
S.~C. Miller and W.~F. Love, {\em Tables of Irreducible Representations of
  Space Groups and Co-Representations of Magneticd Space Groups} (Pruett,
  Bolder, 1967).

\bibitem{stokes88:spcgrp}
H.~T. Stokes and D.~M. Hatch, {\em Isotropy Subgroups of the 230
  Crystallographic Space Groups} (World Scientific, Singapore, 1988).

\bibitem{birch78:eos}
F. Birch, J.~Geophys. Res. {\bf 83},  1257  (1978).

\bibitem{mehl94:fpcij}
M.~J. Mehl, B.~M. Klein, and D.~A. Papaconstantopoulos,  in {\em Intermetallic
  Compounds - Principles and Practice}, edited by J. Westbrook and R. Fleischer
  (John Wiley and Sons, London, 1994), Vol.~1, Chap.~9, pp.\ 195--210.

\end{thebibliography}
\end{document}